\documentclass[eps,iop,superscriptaddress,amssymb,eqsecnum,showpacs,showkeyes,secnumarabic,graphics,floatfix,nofootinbib,tightenlines,longbibliography,notitlepage,11pt]{revtex4-1}

\usepackage{graphicx}
\usepackage{bm}
\usepackage{dcolumn}
\usepackage{amsmath}
\usepackage{cancel}
\usepackage{mathtools}
\usepackage[utf8]{inputenc}
\usepackage[breaklinks=true,colorlinks=true,
linkcolor=blue,urlcolor=blue,citecolor=cyan,
bookmarks=true,bookmarksopenlevel=2]{hyperref}

\def \beq  {\begin{equation}}
\def \eeq  {\end{equation}}
\def \ber  {\begin{eqnarray}}
\def \eer  {\end{eqnarray}}

\begin{document}
\newcommand{\newc}{\newcommand}

\newc{\be}{\begin{equation}}
\newc{\ee}{\end{equation}}
\newc{\ba}{\begin{eqnarray}}
\newc{\ea}{\end{eqnarray}}
\newc{\bea}{\begin{eqnarray*}}
\newc{\eea}{\end{eqnarray*}}
\newc{\D}{\partial}
\newc{\ie}{{\it i.e.}}
\newc{\eg}{{\it e.g.}}
\newc{\etc}{{\it etc.}}
\newc{\etal}{{\it et al.}}
\newcommand{\nn}{\nonumber}
\newc{\ra}{\Rightarrow}
\title{A general realistic treatment of the disk paradox}
\author{George Pantazis}\email{ph06767@cc.uoi.gr}
\author{Leandros Perivolaropoulos}\email{leandros@uoi.gr}
\affiliation{Department of Physics, University of Ioannina, GR-45110, Ioannina, Greece}

\date {\today}

\begin{abstract}
Mechanical angular momentum is not conserved in systems involving electromagnetic fields with non-zero electromagnetic field angular momentum. Conservation is restored only if the total (mechanical and field) angular momentum is considered. Previous studies have investigated this effect, known as ``Feynman's Electromagnetic Paradox" or simply ``Disk Paradox" in the context of idealized systems (infinite or infinitesimal solenoids and charged cylinders \etc). In the present analysis we generalize previous studies by considering more realistic systems with finite components and demonstrating explicitly the conservation of the total angular momentum. This is achieved by expressing both the mechanical and the field angular momentum in terms of charges and magnetic field fluxes through various system components. Using this general expression and the closure of magnetic field lines, we demonstrate  explicitly the conservation of total angular momentum in both idealized and realistic systems (finite solenoid concentric with two charged cylinders, much longer than the solenoid) taking all the relevant terms into account including the flux outside of the solenoid. This formalism has the potential to facilitate a simpler and more accurate demonstration of total angular momentum conservation in undergraduate physics electromagnetism laboratories.
\end{abstract}
\maketitle

\section{Introduction}
\label{sec:Introduction}

The conversion of electromagnetic field angular momentum into mechanical angular momentum may be demonstrated using a variety of simple experimental setups, where the time variation of the overlapping electric and magnetic fields leads to variation of total mechanical angular momentum of charged objects constituting the system. This conversion which may be viewed as non-conservation of mechanical angular momentum is known as {\it{Feynman's Electromagnetic Paradox}} (FEP) \cite{Feynman:1964:FLP}.

\begin{figure}[t!]
\centering
\includegraphics[width=5cm,height=8cm]{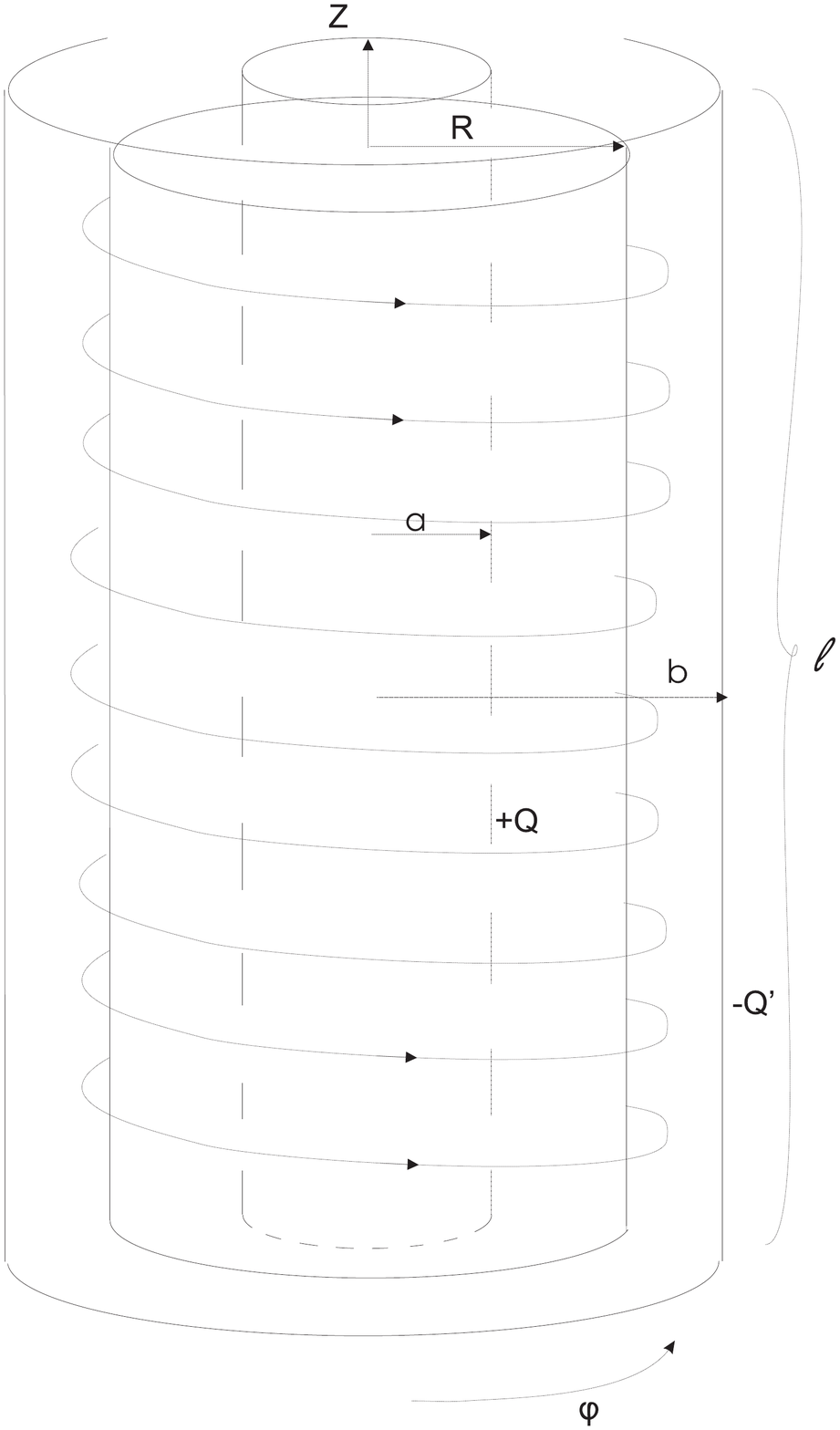}
\caption{The physical system considered consists of two long but finite charged cylindrical shells with a concentric solenoid in the space between them. The charges $Q$ and $Q^\prime$ refer to the region of the solenoid of length $l$ and the linear densities of the cylinders are $\lambda$ and $\lambda^\prime$ respectively (and therefore $Q= \lambda l$ and $Q^\prime = \lambda^\prime l$). The length of the cylinders is assumed to be much longer than the solenoid, even though this is not completely apparent in the figure.}
\label{fig:Figure_1}
\end{figure}

The quantitative theoretical investigation of this phenomenon has been performed in the context of various simple systems \cite{1967AmJPh..35..153P, 1966AmJPh..34..772R, 1985AmJPh..53...15R, 1980AmJPh..48...83C, 1985AmJPh..53..495B, :/content/aapt/journal/ajp/52/8/10.1119/1.13568, 1983AmJPh..51..213L, 1986AmJPh..54..949M, :/content/aapt/journal/ajp/56/5/10.1119/1.15592, 1989AmJPh..57..558G, :/content/aapt/journal/ajp/62/1/10.1119/1.17738, :/content/aapt/journal/ajp/52/8/10.1119/1.13853, Griffiths, Belcher}. Most of these analyses study either infinitesimal \cite{1985AmJPh..53..495B, 1986AmJPh..54..949M, :/content/aapt/journal/ajp/62/1/10.1119/1.17738}
or infinite \cite{1966AmJPh..34..772R, 1985AmJPh..53...15R, :/content/aapt/journal/ajp/52/8/10.1119/1.13853, Griffiths, Belcher} solenoids for the production of the required magnetic field. Refs.~\cite{:/content/aapt/journal/ajp/52/8/10.1119/1.13853, Griffiths} (see also \cite{1966AmJPh..34..772R})  consider a system of two infinite concentric cylindrical shells uniformly charged with opposite total charges $(Q,-Q)$. An infinite solenoid with initial current density $i$ is placed concentric with the shells in the region between them (Fig.~\ref{fig:Figure_1}).

The momentum density \cite{Griffiths} of an electromagnetic field is
\be
\vec{p}_{\text{em}}=\epsilon_0\mu_0 \ \vec{S} =\epsilon_0 \  \vec{E}\times\vec{B}, 
\ee
$\vec{S}$ is the Poynting vector. Hence, the angular momentum density of an electromagnetic field is defined as
\be \vec{\mathcal{L}}_\text{em}=\vec{r}\times\vec{p}_\text{em}
=\epsilon_0 \ \vec{r}\times\left(\vec{E}\times\vec{B}\right). \label{ang_mom_field} \ee

As discussed in \cite{:/content/aapt/journal/ajp/52/8/10.1119/1.13853, Griffiths}, when the current of the solenoid is reduced to zero the field's angular momentum \eqref{ang_mom_field} goes to zero and gets transformed into mechanical angular momentum of the shells. This is verified in \cite{:/content/aapt/journal/ajp/52/8/10.1119/1.13853, Griffiths} by comparing the total initial field's angular momentum with the mechanical angular momentum generated by the torque of the Faraday induced electric field on the charged shells as the solenoid magnetic field decreases to zero.

In these studies the analysis is simple and demonstrates the conservation of total angular momentum in the simplest possible manner. However, their analysis ignores two important effects which would be present in any realistic lab system:

\begin{itemize}
\item[1.] It ignores the magnetic field generated by the rotating charged cylinders and thus assumes that the field's angular momentum vanishes with the solenoid current. However, see problem $8.13$ of Ref.~\citep{Griffiths}.

\item[2.] It assumes an infinite system where there is no return magnetic flux out of the solenoid and the cylinders. This leads to non-conservation of total angular momentum if the analysis
is naively generalized to the case where the charges of the two cylinders are not opposite but arbitrary (unshielded electric field). In that case, the return flux of the long but finite solenoid needs to be taken into account as pointed out in \cite{Belcher}.
\end{itemize}

The first assumption does not spoil the conservation of total angular momentum but clearly affects the accuracy of the analysis and its applicability in laboratory setups. The resolution of the second problem requires one to assume a finite system and take into account the return flux of the magnetic field lines of the solenoid and the rotating cylinders in the outside region. When the cylinder charges are opposite, this return flux can be neglected and has no angular momentum because there is (almost) no  electric field in the outside region (it is shielded by the outer cylinder). For arbitrary cylinder charges however, the return flux needs to be taken into account \cite{Belcher} in order to secure angular momentum conservation.

The demonstration of angular momentum conservation in an undergraduate physics laboratory requires of both the total mechanical angular momentum and the field angular momentum, before and after changing the electromagnetic field components throughout space. In view of the complexity of the magnetic field in finite realistic systems like solenoids, this estimate may become a complex non-trivial task if all relevant contributions are to be included. In our analysis we demonstrate that the precise knowledge of the magnetic fields is not necessary to demonstrate total angular momentum conservation. Instead, the magnetic fluxes and the charges of the system components are enough for evaluating both the electromagnetic field and the mechanical angular momentum generated by electromagnetic field changes. The closure of magnetic field lines, provides connections between the magnetic fluxes of various system regions and thus simplifies the estimate of these fluxes through outer regions of the system. Therefore, this formalism has the potential to facilitate a simpler and more accurate demonstration of total angular momentum conservation in undergraduate physics electromagnetism laboratories.

In particular, in the present study we extend the analyses of \cite{:/content/aapt/journal/ajp/52/8/10.1119/1.13853, Griffiths} and demonstrate total angular momentum conservation in the system of Fig.~\ref{fig:Figure_1} for a long but finite system and arbitrary cylinder charges $Q$ and $-Q^\prime$. We make no assumption about the length of the solenoid but we assume that the length of the cylinders is much longer than the length of the solenoid so that the electric field in the region of the solenoid has a simple form. Our analytical treatment takes advantage of the closure of the magnetic field lines, allowing us the trivial evaluation of the magnetic field flux outside the system (it is equal in magnitude to the inside flux). Thus, it becomes simple to verify angular momentum conservation once the mechanical and field angular momenta are expressed in terms of magnetic field fluxes. Such expressions simplify dramatically the analysis and allow its applicability to very general setups where the magnetic field form may be complicated (especially outside the solenoid/cylinders). In this study, we take into account the magnetic field of the solenoid, the rotating cylinders and their corresponding return flux. Thus, our analysis is more realistic and should be useful in any attempt to demonstrate the effect in a lab. As a special case we consider the known analytic form of a magnetic field of a finite solenoid\cite{2015arXiv150705075L}.

The structure of this paper is the following: In Section~\ref{sec:Section 2} we express the total field and mechanical angular momenta for a generalized magnetic field in terms of magnetic fluxes and use these expressions to demonstrate conservation of total angular momentum in a general setup independent of the detailed form of the magnetic field. In Section~\ref{sec:Section 3} we consider the special case in which the magnetic field is approximatelly homogeneous corresponding to a long solenoid, while in Section~\ref{sec:Section 4} we apply the general formalism to the more realistic case of a finite arbitrary size solenoid. Finally, we conclude and discuss our results in~\ref{sec:Conclusions}.

\section{Generalized Analysis Independent of Magnetic Field}
\label{sec:Section 2}

We assume a general axially symmetric magnetic field of the form
\be
\vec{B}=B_\rho^{(s)}(\rho, z)\hat{\rho}+B_z^{(s)}(\rho, z)\hat{z}, \label{gen_magn_field} 
\ee
where $^{(s)}$ stands for the source of the magnetic field ($sol$ for solenoid, $a$ for the rotating cylinder $a$ and $b$ for the rotating cylinder $b$) of Fig.~\ref{fig:Figure_1}. We use cylindrical coordinates and we define
\be
\vec{r}=\rho {\hat \rho}+z {\hat z}.
\ee
Due to the azimuthal symmetry of the system, it can be shown that, disregarding a small current flow in the $z$ direction, we have $B_\phi^{(s)}=0$ (\eg \ see \cite{0143-0807-37-2-025203}) and also that $B_\rho^{(s)}$ and $B_z^{(s)}$ do not depend on $\phi$.

The initial state of the system of Fig.~\ref{fig:Figure_1} is such that for $a\leq \rho \leq b$, \ie \ in the range close to the solenoid (\ie \ $z \in [-l/2,l/2]$) there is an electric field
\be \vec{E}_< =\frac{Q}{2\pi\epsilon_0 l \rho}\Theta(\rho -a)\Theta(b-\rho)\hat{\rho} \equiv E_< \hat{\rho} \label{elect_field_in} \ee
while for $\rho>b$ we have
\be
\vec{E}_> = \frac{Q-Q^\prime}{2\pi\epsilon_0 l \rho}\Theta(\rho -b)\hat{\rho} \equiv E_> \hat{\rho}, \label{elect_field_out} 
\ee
according to Gauss's law (where $l$ is the length of the solenoid where the charges $Q$ and $Q^\prime$ are contained, while the length of the cylinders is $L\gg l$). We assume that the cylinders are much longer than the solenoid so that we can ignore the edge effects of the electric field. The solenoid extends for $z \in [-l/2,l/2]$ and we focus on the angular momenta (field and mechanical) in this region.

We thus require the validity of equations \eqref{elect_field_in} and \eqref{elect_field_out} in the region close to the solenoid. Far away from the solenoid where the magnetic field is negligible, the form of the electric field is irrelevant since there is no field angular momentum or induced mechanical angular momentum there.

Similarly, the magnetic field inside the solenoid\footnote{From now on, when we refer to the magnetic field inside and outside the solenoid we use the notation $\prec$ and $\succ$ respectively, while for the electric field we use $<$ in the region $a \leq \rho \leq b$ and $>$ in the region $\rho > b$.}, \ie \ $\rho\leq R$ is
\be
\vec{B}_{sol_\prec} (\rho, z) = B_\rho^{sol_\prec} \Theta(R-\rho) \ \hat{\rho} + B_z^{sol_\prec} \Theta(R-\rho) \ \hat{z},
\label{magn_field_gen_in}
\ee
while outside the solenoid is
\be \vec{B}_{sol_\succ} (\rho, z) = B_\rho^{sol_\succ} \Theta(\rho - R) \ \hat{\rho} +B_z^{sol_\succ} \Theta(\rho - R) \ \hat{z},
\label{magn_field_gen_out}
\ee
where $R$ is the solenoid's radius.

We now demonstrate that the utilization of closed field lines is powerful enough to lead to expression of both the mechanical and field angular momenta in terms of cylinder charges and magnetic fluxes only, with no reference to the detailed form of the magnetic field. Thus, angular momentum conservation is demonstrated for {\it any} form of the magnetic field of the solenoid.

\begin{figure}[h!]
\centering
\includegraphics[height=8cm]{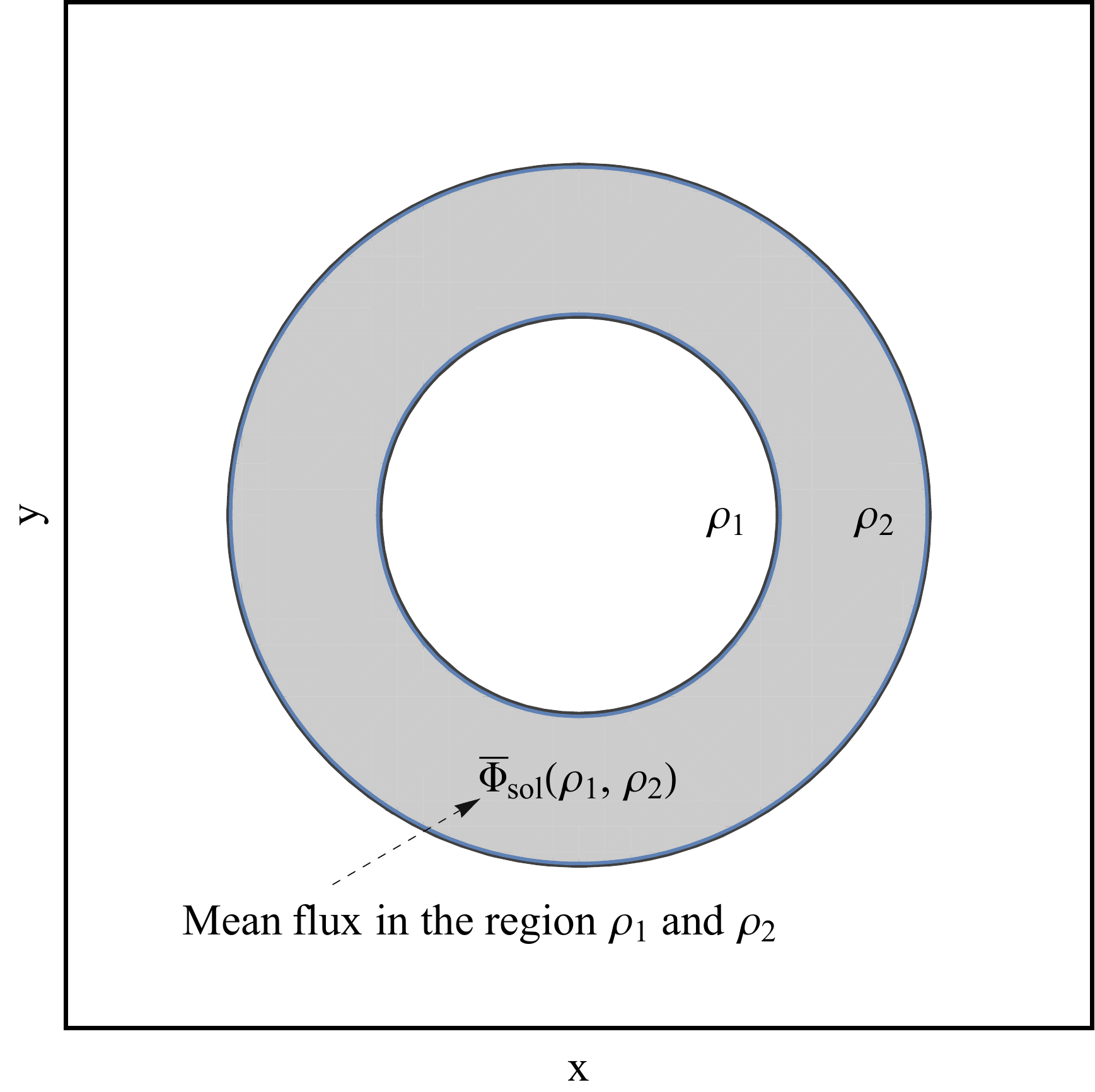}
\caption{The shadowed region in which the mean flux is calculated.}
\label{fig:Figure_2}
\end{figure}

Our starting point is the definition of the mean magnetic flux through cylindrical spatial region along the symmetry axis of the system. We define this mean magnetic flux as
\be
\bar{\Phi}_{(s)}(\rho_1, \rho_2)\equiv \frac{1}{l}\int_{-l/2}^{l/2}dz \ \Phi_{(s)}(\rho_1,\rho_2)= \frac{2\pi}{l}\int_{-l/2}^{l/2} dz^\prime \int_{\rho_1}^{\rho_2} d\rho^\prime \ \rho^\prime \ B_z^{(s)} (\rho^\prime, z^\prime)
, \label{fluxdef} 
\ee
where the index $(s)$ refers to the source of the magnetic field, which is an even function of $z$. With this definition, it is straightforward to evaluate the angular momentum of each source (solenoid or cylinder), as well as the corresponding mechanical momentum it induces on the cylinders, in terms of just cylinder charges ($Q$ and $Q^\prime$) and fluxes of the form \eqref{fluxdef}. We use the notation that $\Phi_{(s)}^\prec(\rho_1, \rho_2)$ is the magnetic flux due to the magnetic field inside the source and $\Phi_{(s)}^\succ(\rho_1, \rho_2)$ the corresponding magnetic flux due to the magnetic field lines outside the source (return field lines). In Figure \ref{fig:Figure_2} we show the region used in the calculation of the mean flux of equation \eqref{fluxdef}.

Initially we assume that there is no mechanical angular momentum and that there is only field angular momentum due to the magnetic field of the solenoid and the electric field of the cylinders. Using equations \eqref{ang_mom_field}, \eqref{gen_magn_field}, \eqref{elect_field_in} and \eqref{elect_field_out} we may write for the field's angular momentum density
\be
\begin{split}
\vec{\mathcal{L}}_{sol} &= \epsilon_0 \ \vec{r} \times \left(\vec{E} \times\vec{B}\right)= \\ &=
\epsilon_0 \ {\vec r} \times \left[ E_<{\hat \rho} \times \left( B_\rho^\prec(\rho, z) \ {\hat \rho} + B_z^\prec(\rho, z) \ {\hat z} \right) \right. \\  & \left. +
\left(E_< + E_> \right) {\hat \rho} \times \left(B_\rho^\succ(\rho, z) \ {\hat \rho} + B_z^\succ(\rho, z) \ {\hat z} \right) \right]= \\ &=
\left(\rho {\hat \rho}+z {\hat z} \right) \times (-{\hat \phi}) \left[E_< B_z^\prec(\rho, z)+ \left( E_< + E_> \right) B_z^\succ(\rho, z) \right] \ra \nn
\end{split}
\ee
\begin{equation}
\vec{\mathcal{L}}_{sol} = \epsilon_0 \left( z \hat{\rho} - \rho\hat{z} \right) \left[ E_< B_z^\prec + \left( E_< + E_> \right) B_z^\succ \right],
\label{field_ang_mom_s}
\end{equation}
where we have assumed a right-handed coordinate system. Using equations \eqref{elect_field_in}, \eqref{elect_field_out} and \eqref{fluxdef} we may express the total field angular momentum\footnote{In what follows we label total field angular momentum as $\mathcal{L}^\text{tot}$, while the field angular momentum density as $\mathcal{L}$. In addition, the mechanical angular momentum is refered as $L$.} of the system in terms of magnetic fluxes as
\be
\vec{\mathcal{L}}_{sol}^{tot} = \left[\frac{Q}{2\pi} \left(\bar{\Phi}_{sol}^\prec (a,R)+\bar{\Phi}_{sol}^\succ(R,b)\right)+\frac{Q-Q^\prime}{2\pi} \bar{\Phi}_{sol}^\succ(b,\infty) \right](-{\hat z}), 
\label{l_field_s}
\ee
where $\bar{\Phi}_{sol}^\succ(b,\infty)$ and $\bar{\Phi}_{sol}^\succ(R,b)$ are the outside return fluxes of the solenoid (beyond radius $R$) which are assumed to satisfy
\be
\bar{\Phi}_{sol}^\succ(b,\infty)+\bar{\Phi}_{sol}^\succ(R,b)= -\bar{\Phi}_{sol}^\prec(0,R) \label{clos_f_lines_s} 
\ee
due to the closure of the magnetic field lines. In far regions where the magnetic field is negligible (\ie \ $z \gg l$), the exact form of the electric field is irrelevant since there is no field angular momentum in these regions. Thus for the validity of equation \eqref{l_field_s} we  only assume that the cylinders are much longer than the solenoid and thus equations \eqref{elect_field_in} and \eqref{elect_field_out} are applicable in the region of the solenoid. Equation \eqref{clos_f_lines_s} is valid only for finite solenoids and its validity may be demonstrated by considering the integral form of Gauss law for magnetism. For a fixed $z$, we consider  a closed surface consisting of an plane (perpendicular to the solenoid) closed at infinity (where the flux is assumed zero for a finite solenoid). Then Gauss law of magnetism implies that the total magnetic flux across the plane perpendicular to the solenoid vanishes. The minus sign enters due to the reversal of the magnetic field which also reverses the direction of the field angular momentum density. Notice that the radial part of the magnetic field contributes neither to the field's total angular momentum nor to the total mechanical momentum of the cylinders and therefore using equation \eqref{fluxdef} the radial part of \eqref{field_ang_mom_s} drops out (as an integral of an odd function). Equations \eqref{l_field_s} and \eqref{clos_f_lines_s} lead to
\be
\vec{\mathcal{L}}^{tot}_{sol}=\frac{{\hat z}}{2\pi}\left[Q \  \bar{\Phi}_{sol}^\prec(0,a) - Q^\prime \left(\bar{\Phi}_{sol}^\prec(0,R) +\bar{\Phi}^\succ_{sol}(R,b)\right) \right]=\frac{{\hat z}}{2\pi}\left[Q \  \bar{\Phi}_{sol}^\prec(0,a) + Q^\prime \bar{\Phi}^\succ_{sol}(b,\infty) \right]. \label{lfield_s} 
\ee
We now assume that current in the solenoid drops converting field angular momentum to mechanical angular momentum. Using Faraday's law we obtain
\be
\oint \vec{E}^F \cdot d\vec{l}=- \frac{d}{dt} \left(\Phi_{sol}+\Phi_a+\Phi_b \right),
\ee
where $\Phi_{sol}$ is the flux due to the solenoid at a particular $z$ and $\Phi_a$, $\Phi_b$ are the magnetic fluxes due to the rotating charged cylinders. The index $F$ implies that this is the Faraday induced electric field. Hence, for the cylinder $a$ it is
\be
\oint \vec{E}^F \cdot d\vec{l}=\int_0^{2\pi} \vec{E}^F \cdot {\hat \phi} \ \rho \ d\phi=-\frac{d \Phi_{sol}}{dt} \ra \vec{E}_a^{sol,F} =-\frac{ \frac{d}{dt}\Phi_{sol}(0,a)}{ 2\pi \rho}{\hat \phi},
\ee
while the corresponding torque for the inner cylinder $a$ due to the solenoid is
\be
\vec{N}_a^{sol}= Q \ \vec{r}\times \vec{E}_a^{sol,F}=\left(\rho {\hat \rho}+z {\hat z} \right)\times ({-\hat \phi}) \frac{Q}{2\pi \rho}\frac{d}{dt}\Phi_{sol}(0,a)= \frac{Qz}{2\pi\rho}\frac{d}{dt}\Phi_{sol}(0,a){\hat \rho} + \frac{Q}{2\pi}\frac{d}{dt}\Phi_{sol}(0,a) \left(-{\hat z}\right).
\ee
By integrating the torque over time and taking the mean magnetic flux as in equation \eqref{fluxdef}, the radial part similarly drops out as in equation \eqref{field_ang_mom_s} and we write the mechanical angular momentum of cylinder $a$ due to the solenoid which is
\be
{\vec L}_a^{sol}= \frac{Q}{2\pi} \bar{\Phi}_{sol}(0,a) {\hat z}, \label{lmech_s_a} \ee
while, similarly for the outer cylinder $b$ we have
\be
{\vec L}_b^{sol}= \frac{-Q^\prime}{2\pi}\left( \bar{\Phi}_{sol}^\prec(0,R)+\bar{\Phi}^\succ_{sol}(R,b)\right) {\hat z}= \frac{Q^\prime}{2\pi}\bar{\Phi}^\succ_{sol}(b,\infty) {\hat z}\label{lmech_s_b} 
\ee
From equations \eqref{lfield_s}$-$\eqref{lmech_s_b} it becomes clear that
\be
\Delta\left({\vec L}^{sol}+\vec{\mathcal{L}}^{tot}_{sol}\right)=\left({\vec L}_a^{sol}+{\vec L}_b^{sol}-0\right)+\left(0-\vec{\mathcal{L}}^{tot}_{sol}\right)=0, \label{d_lmech_lfield_s}
\ee
where $\Delta$ denotes the difference ``final minus initial value". Therefore, the total angular momentum induced by the solenoid is conserved independent of the precise form of the magnetic field.

The above analysis, takes into account only the magnetic field of the solenoid. However, the cylinders also induce magnetic field as they rotate and it also contributes to the angular momentum. In order to take it into account we perform a similar analysis for the total angular momentum induced by each one of the cylinders. In particular, for the field angular momentum of the cylinder $a$ we find
\be
\vec{\mathcal{L}}^{tot}_{a}={-\hat z}\left[\frac{(Q-Q^\prime)}{2\pi} \ \bar{\Phi}^\succ_a(b, \infty)+\frac{Q}{2\pi} \ \bar{\Phi}^\succ_a(a,b) \right]={\hat z} \left[\frac{(Q-Q^\prime)}{2\pi} \ \bar{\Phi}^\prec_a(0,a)-\frac{Q^\prime}{2\pi} \ \bar{\Phi}^\succ_a(a,b) \right], 
\label{lfield_a} 
\ee
where $\bar{\Phi}^\succ_a(b, \infty)+\bar{\Phi}^\succ_a(a,b)= -\bar{\Phi}^\prec_a(0,a)$ are the outside fluxes of cylinder $a$ which their sum is opposite of the inside flux $\bar{\Phi}^\prec_a(0,a)$. The mechanical angular momenta ${\vec L}_a^a$, ${\vec L}_b^a$ induced on the cylinders $a$ and $b$ due to the increasing magnetic field of cylinder $a$ are easily found to be
\ba 
{\vec L}_a^a &= \frac{Q}{2\pi} \bar{\Phi}^\prec_a(0,a) (-{\hat z}), \label{lmech_a_a} \\ {\vec L}_b^a &= \frac{-Q^\prime}{2\pi} \bar{\Phi}^\succ_a(b, \infty) (-{\hat z}) &=\frac{-Q^\prime}{2\pi} \left(\bar{\Phi}^\prec_a(0,a)+\bar{\Phi}^\succ_a(a,b)\right) {\hat z}. \label{lmech_a_b} 
\ea
Using equations \eqref{lfield_a}$-$\eqref{lmech_a_b} we find the conservation of the total angular momentum due to cylinder $a$ as
\be
\Delta\left({\vec L}^a+\vec{\mathcal{L}}^{tot}_{a}\right)=\left({\vec L}_a^a+{\vec L}_b^a - 0\right)+\left(\vec{\mathcal{L}}^{tot}_{a}-0\right)=0. \label{d_lmech_lfield_a} 
\ee
At this point, it is important to mention that the electric field in the rotating case has the same radial component as the non-rotating one. In addition it has also a $\phi$ component due to Faraday's law, which is absent in the non-rotating case. The latter contributes only to a $\rho$ component and therefore we have a $z$ component of the angular momentum in equation~\eqref{lfield_a}.

Finally, for the field and mechanical angular momentum induced by cylinder $b$ we find
\be 
\vec{\mathcal{L}}^{tot}_{b}=\left[\frac{Q}{2\pi}\left( \bar{\Phi}^\prec_b(a,b)\right)+\frac{Q-Q^\prime}{2\pi}\bar{\Phi}_b^\succ(b, \infty)\right](-{\hat z}), 
\label{l_field_b} 
\ee
where $\bar{\Phi}_b^\succ(b, \infty)=-\bar{\Phi}^\prec_b(0, b)$ is the outside flux due to the rotating cylinder $b$ which due to field line closure is opposite to the inside flux. We thus obtain
\be 
\vec{\mathcal{L}}^{tot}_{b}=\frac{{\hat z}}{2\pi}\left[Q \ \bar{\Phi}^\prec_b(0,a)-Q^\prime \ \bar{\Phi}^\succ_b(b, \infty) \right]. 
\label{lfield_b} 
\ee
For the mechanical angular momentum induced on the cylinders due to cylinder $b$ we find
\be
{\vec L}_a^b= \frac{Q}{2\pi} \bar{\Phi}_b^\prec(0,a) (-{\hat z}) \label{lmech_a_b}
\ee
for cylinder $a$ and
\be
{\vec L}_b^b= \frac{-Q^\prime}{2\pi} \bar{\Phi}^\succ_b(b, \infty) (-{\hat z}) 
\label{lmech_b_b}
\ee
for cylinder $b$. Using equations \eqref{lfield_b}$-$\eqref{lmech_b_b} we find the conservation of total angular momentum due to cylinder $b$ as
\be
\Delta\left({\vec L}^b+\vec{\mathcal{L}}^{tot}_{b}\right) =\left({\vec L}_a^b+{\vec L}_b^b - 0\right) +\left(\vec{\mathcal{L}}^{tot}_{b}-0\right)=0. \label{d_lmech_lfield_b}
\ee
Summation of equations \eqref{d_lmech_lfield_s}, \eqref{d_lmech_lfield_a} and \eqref{d_lmech_lfield_b} demonstrates conservation of total angular momentum of the system, induced by the solenoid, cylinder $a$ and cylinder $b$. We have
\ba
{\vec L}_a&+&{\vec L}_b={\vec L}_a^{sol}+{\vec L}_a^a+{\vec L}_a^b+{\vec L}_b^{sol}+{\vec L}_b^a+{\vec L}_b^b= \nn \\
&=& \frac{{\hat z} Q}{2\pi} \left[\bar{\Phi}^\prec_{sol}(0,a)-\bar{\Phi}^\prec_a(0,a)-\bar{\Phi}^\prec_b(0,a)
\right]+\frac{{\hat z} Q^\prime}{2\pi} \left[\bar{\Phi}^\succ_{sol}(b, \infty)-\bar{\Phi}^\succ_a(b, \infty)+\bar{\Phi}^\succ_b(b, \infty)\right] 
\label{tot_lmech}
\ea
and
\ba
\vec{\mathcal{L}}^{tot}_{i}&-&\vec{\mathcal{L}}^{tot}_{f}=\vec{\mathcal{L}}^{tot}_{sol}-\vec{\mathcal{L}}^{tot}_{a}-\vec{\mathcal{L}}^{tot}_{b}= \nn \\
&=& \frac{{\hat z} Q}{2\pi} \left[\bar{\Phi}^\prec_{sol}(0,a)-\bar{\Phi}^\prec_a(0,a)-\bar{\Phi}^\prec_b(0,a)
\right]+\frac{{\hat z} Q^\prime}{2\pi} \left[\bar{\Phi}^\succ_{sol}(b, \infty)-\bar{\Phi}^\succ_a(b, \infty)+\bar{\Phi}^\succ_b(b, \infty)\right]. 
\label{tot_lfield}
\ea
Clearly equations \eqref{tot_lmech} and \eqref{tot_lfield} satisfy total angular momentum conservation namely 
\be
\vec{\mathcal{L}}^{tot}_{i}-\vec{\mathcal{L}}^{tot}_{f}={\vec L}_a+{\vec L}_b. \label{l_mech_field_cons}
\ee

\section{Application I: System with homogeneous magnetic field}
\label{sec:Section 3}

In the special case of a long solenoid the system is well simplified since the magnetic field is homogeneous and the flux terms of equations \eqref{tot_lmech} and \eqref{tot_lfield} may be analytically calculated. We assume the same electric field equations \eqref{elect_field_in} and \eqref{elect_field_out} and we review the form of the magnetic field inside the solenoid and the rotating cylinders in order to obtain the corresponding inside and outside magnetic fluxes for equations \eqref{tot_lmech} and \eqref{tot_lfield}. 

For $\rho \leq R$, where $R$ is the solenoid's radius  (Fig.~\ref{fig:Figure_1}), the initial magnetic field is
\be
\vec{B}_{sol_\prec} = \mu_0 i \ {\hat z},
\label{b_s_in}
\ee
where $i= N I$ is the current density and $N$ the number of turns per unit length. 

The magnetic field due to the cylinder $b$ can be calculated using Amp\'ere law. Thus, considering an orthogonal Amperian contour of length  $l$, we have:
\be
\begin{aligned}
 \oint\vec{B}_{b_{\prec}} \cdot d\vec{l} = \mu_{0}i_{enc}=\mu_{0}\frac{|-Q^\prime|}{T_b} \ra {B}_{b_{\prec}} = \mu_{0} \frac{|-Q^\prime|}{2\pi{l}}\omega_b= \mu_{0}|\sigma_b| \ b \ \omega_b,
\end{aligned}
\ee
where $B_{b_{\prec}}$ is the magnitude of the magnetic field induced by the outer cylinder at $\rho <b$, $T_b$ is the cylinder rotation period, $\omega_b$ is the corresponding angular velocity and \be |-Q^\prime|=|\sigma_b|2\pi b l,\ee where $\sigma_b$ is the $b$ cylinder surface charge density. The direction of the final magnetic field may be obtained using  Lenz's law as
\be
\vec{B}_{b_f{\prec}}=\mu_{0}|\sigma_{b}|b\omega_{b_f}\hat{z},
\label{magn_f_final_lenz_b}
\ee
where the additional index $f$ stands for the final value of the field (after the current has gone to zero). 

Similar results may be shown for cylinder $a$. In particular
\be
\vec{B}_{a_f{\prec}}=\mu_{0}|\sigma_{a}|a\omega_{a_f} {\hat z}
 \label{magn_f_final_lenz_a}
\ee

Using now equations (\ref{b_s_in}),(\ref{magn_f_final_lenz_a}) and  (\ref{magn_f_final_lenz_b}) in equations (\ref{tot_lfield}) and (\ref{tot_lmech}) we have

\be
\begin{split}
\vec{\mathcal{L}}^{tot}_{i} -\vec{\mathcal{L}}^{tot}_{f} &= {\hat z}\frac{Q}{2\pi}\left[ \left(\mu_0 i \right)\pi a^2-\left( \mu_0 \left| \sigma_a \right| a \omega_{a_f} \right)\pi a^2-\left(\mu_0 \left|\sigma_b\right| b \omega_{b_f}\right)\pi a^2 \right] \\ &+ {\hat z}\frac{Q^\prime}{2\pi}\left[\bar{\Phi}^\succ_{sol}(b, \infty)-\bar{\Phi}^\succ_a(b, \infty)+\bar{\Phi}^\succ_b(b, \infty)\right]
\end{split}
\label{dam-homog}
\ee
and similarly for $\vec{L}_a+\vec{L}_b $.
This equation may be simplified further if we assume that the field lines of the cylinders and the solenoid close beyond the cylinder $b$ (ignore return fluxes in the range between $a$ and $b$). In this case we have 
\ba
\bar{\Phi}^\succ_{sol}(b, \infty)&=&-\bar{\Phi}^\prec_{sol}(0, R)=-\left(\mu_0 i \right)\pi R^2 \\
\bar{\Phi}^\succ_a(b, \infty)&=&-\bar{\Phi}^\prec_a(0, a)=-\left( \mu_0 \left| \sigma_a \right| a \omega_{a_f} \right)\pi a^2 \\
\bar{\Phi}^\succ_b(b, \infty)&=&-\bar{\Phi}^\prec_b(0, b)=-\left( \mu_0 \left| \sigma_b \right| b \omega_{b_f} \right)\pi b^2
\ea
In this approximation equation (\ref{dam-homog}) gets simplified as
\be
\vec{\mathcal{L}}^{tot}_{i} -\vec{\mathcal{L}}^{tot}_{f} = {\hat z}\frac{\mu_0 \left| \sigma_b\right| \omega_{b_f} b}{2} \left(Q^\prime b^2 -Q a^2\right)+{\hat z}\frac{\mu_0 \left| \sigma_a\right| \omega_{a_f} a^3}{2} \left(Q^\prime -Q\right)+{\hat z}\frac{\mu_0 i}{2}\left(Q a^2-Q^\prime R^2\right).
\label{hom-field-am}
\ee
and similarly for $\vec{L}_a+\vec{L}_b $, \ie
\be
\vec{L}_a+\vec{L}_b = {\hat z}\frac{\mu_0 \left| \sigma_b\right| \omega_{b_f} b}{2} \left(Q^\prime b^2 -Q a^2\right)+{\hat z}\frac{\mu_0 \left| \sigma_a\right| \omega_{a_f} a^3}{2} \left(Q^\prime -Q\right)+{\hat z}\frac{\mu_0 i}{2}\left(Q a^2-Q^\prime R^2\right).
\label{hom-mech-am}
\ee

Therefore, we obtain the angular momentum conservation equation \eqref{l_mech_field_cons},  $\vec{\mathcal{L}}^{tot}_{i} -\vec{\mathcal{L}}^{tot}_{f}=\vec{L}_a+\vec{L}_b$ in terms of simple analytic expression for the fluxes. This simplicity is not retained however in a system where the solenoid has arbitrary finite dimension. This case is discussed in the following section.

\section{Application II: System with realistic dimensions}
\label{sec:Section 4}

In this section we discuss the case for a realistic solenoid of arbitrary size placed between two much longer cylinders. We demonstrate the calculation of the mean fluxes that appear in equations \eqref{tot_lmech} and \eqref{tot_lfield}.
 The electric field stays the same as defined in equations \eqref{elect_field_in} and \eqref{elect_field_out} and the magnetic field is of the form \eqref{gen_magn_field}. It is straightforward to show \cite{2010AmJPh..78..229D, Callaghan, 2015arXiv150705075L} that the magnetic field of a finite size solenoid of length $l$ and radius $R$ at a random point $(\rho,z)$ has components of the form
\ba
B^{(s)}_\rho (\rho, z) &=& -\frac{\mu_0 i R}{2 \pi} \int_0^\pi d\alpha \left[ \frac{\cos\alpha}{ \sqrt{ \xi^2 + \rho^2 + R^2 - 2 R \rho \cos\alpha}}\right]_{\xi_-}^{\xi_+}, \label{b_r_comp} \\
B^{(s)}_z (\rho, z) &=& \frac{\mu_0 i R}{2 \pi} \int_0^\pi d\alpha \left[ \frac{\xi \, (R - \rho \cos\alpha)}{\left( \rho^2 + R^2 - 2 R \rho \cos\alpha \right)\sqrt{ \xi^2 + \rho^2 + R^2 - 2 R \rho \cos\alpha}}\right]_{\xi_-}^{\xi_+}, \label{b_z_comp}
\ea
where, the lengths $\xi_\pm$ are defined by
\be
\xi_\pm = z \pm \frac{l}{2}, \label{eq-xi-pm-defn}
\ee
and $f(\xi)\vert_{\xi_-}^{\xi_+} \equiv f(\xi_+)-f(\xi_-)$.
Equations \eqref{b_r_comp} and \eqref{b_z_comp} may be written in terms of dimensionless variables defined $x=\rho /R$, $\bar{\xi}_\pm=\xi_\pm/R$ and therefore $y=z/R$. Hence, we write
\ba
B^{(s)}_\rho (x, y) &=& -\frac{\mu_0 i}{2 \pi} \int_0^\pi d\alpha \left[ \frac{\cos\alpha}{\sqrt{ \bar{\xi}^2 + x^2 - 2 x \cos\alpha +1}} \right]_{\bar{\xi}_-}^{\bar{\xi}_+}, \label{b_r_comp_x_y} \\
B^{(s)}_z (x, y) &=& \frac{\mu_0 i}{2 \pi} \int_0^\pi d\alpha \left[ \frac{\bar{\xi} \ (1 - x \cos\alpha)}{\left( x^2 - 2 x \cos\alpha +1\right)\sqrt{ \bar{\xi}^2 + x^2 - 2 x \cos\alpha +1}} \right]_{\bar{\xi}_-}^{\bar{\xi}_+}. \label{b_z_comp_x_y}
\ea

It is straightforward to show that equation \eqref{b_r_comp} simplifies to \eqref{b_s_in} for $l\gg R$ \cite{2015arXiv150705075L}.

\begin{figure}[h!]
\centering
\includegraphics[height=8cm]{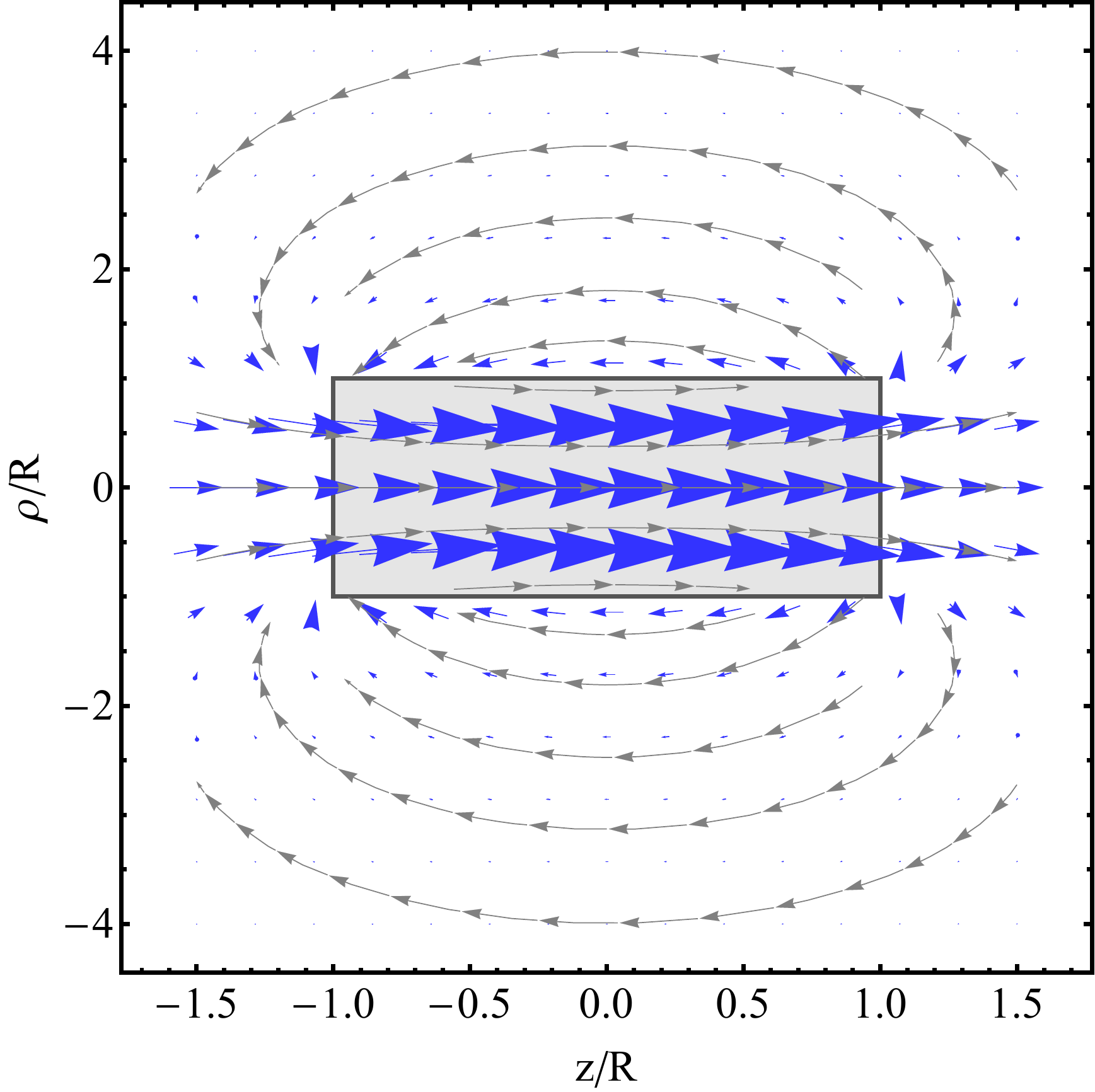}
\caption{Vector and stream magnetic field plot in the region of the solenoid's dimensions (gray region) for $l=2R$.}
\label{fig:Figure_3}
\end{figure}

In Fig.~\ref{fig:Figure_3} we show the vector and the stream magnetic field $\frac{\vec{B}}{\mu_0 i}$ generated for equations \eqref{b_r_comp_x_y} and \eqref{b_z_comp_x_y} and a realistic solenoid (gray region). The magnetic field inside the solenoid is approximately constant despite its physical dimensions.

\begin{figure}[h!]
\centering
\includegraphics[height=6cm]{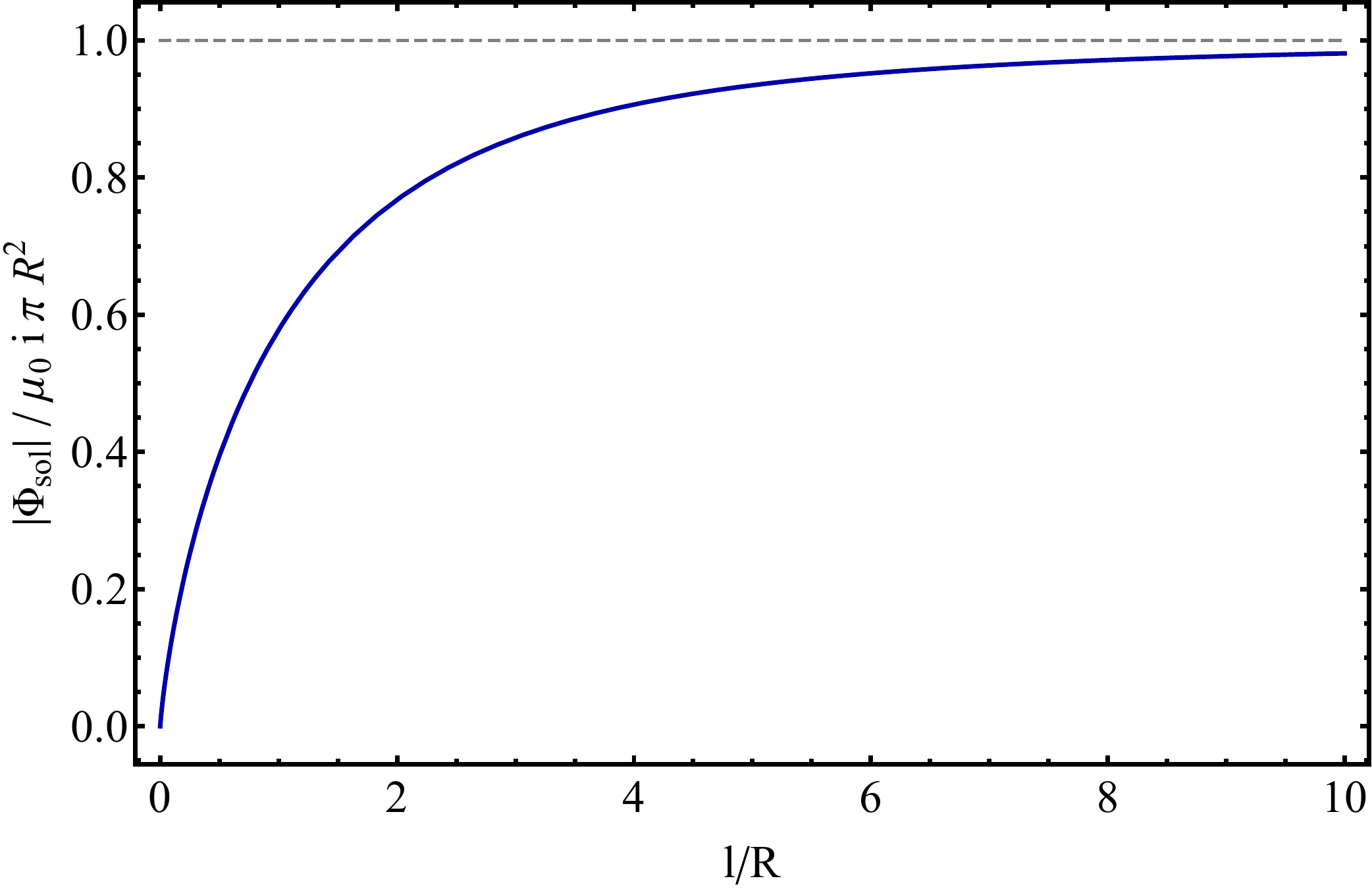}
\caption{Normalized flux $\Phi_{sol}$ at the center of the solenoid ($y=0$) in the dimensionless radial range $[0,1]$ as a function of $l/R$ obtained using equation \eqref{flux_x_y}.}
\label{fig:Figure_4}
\end{figure}

The flux through a dimensionless radial range $[x_1,x_2]$ is obtained using equations \eqref{fluxdef} and \eqref{b_z_comp_x_y} we write
\be
\Phi_{sol} \left(y,x_1,x_2,l/R \right) = \mu_0 i \  R^2  \int_{x_1}^{x_2} dx^\prime \int_0^\pi d\alpha \left[x^\prime \frac{\bar{\xi} \ (1 - x^\prime \cos\alpha)}{\left( x^{\prime 2} - 2 x^\prime \cos\alpha +1\right)\sqrt{ \bar{\xi}^2 + x^{\prime 2} - 2 x^\prime \cos\alpha +1}} \right]_{\bar{\xi}_-}^{\bar{\xi}_+}.
\label{flux_x_y}
\ee
We have verified by numerical integration that 
\be
\Phi_{sol} (0,0,1,\infty) = \mu_0 i \ \pi R^2.
\label{flux_x_y_simp}
\ee
as expected. This is also demonstrated in Fig.~\ref{fig:Figure_4} where we show the normalized  value of the flux at the center of the solenoid as a function of $l/R$. 

The corresponding mean flux as defined in \eqref{fluxdef} is
\be 
\bar{\Phi}_{sol}(x_1, x_2,l/R)= \frac{R}{l} \int_{-\frac{l}{2R}}^{\frac{l}{2R}} dy \; \Phi_{sol}(y,x_1,x_2,l/R)
\label{flux_mean_finite}
\ee

\begin{figure}[h!]
\centering
\includegraphics[width=\textwidth]{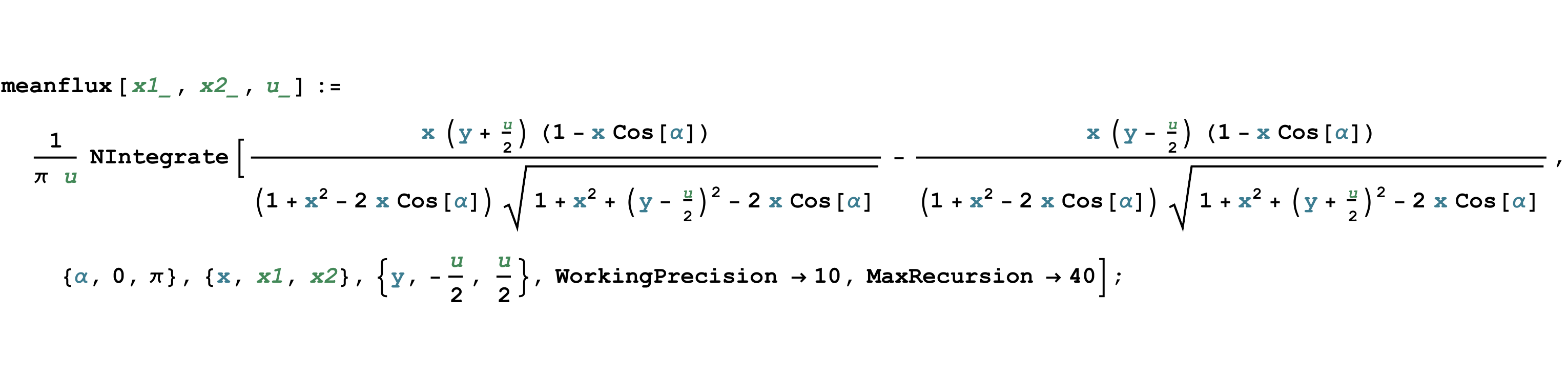}
\caption{Mathematica code for evaluating the RHS of equation \eqref{flux_mean_finite} with $u=\frac{l}{R}$.}
\label{fig:Figure_5}
\end{figure}

We are now in position to evaluate the mean flux terms of equations \eqref{tot_lmech} and \eqref{tot_lfield}. We define the rescaled dimensionless quantities $\frac{a}{R}=\bar{a}$, $\frac{b}{R}=\bar{b}$, $\frac{l}{R}=\bar{l}_{sol}$, $\frac{l_a}{R}=\bar{l}_a$ and $\frac{l_b}{R}=\bar{l}_b$ where $l_a$, $l_b$ are the lengths of the cylinders (much larger than the solenoid length $l\equiv l_{sol}$). Using these definitions we have 
\ba
\bar{\Phi}_{sol}^\prec(0,a)&=&\bar{\Phi}_{sol}(0,\bar{a},\bar{l}_{sol}) \\
\bar{\Phi}_a^\prec(0,a)&=&\bar{\Phi}_{sol} \left(0,1,\frac{\bar{l}_a}{\bar{a}} \right) \\
\bar{\Phi}_b^\prec(0,a)&=&\bar{\Phi}_{sol} \left(0,\frac{\bar{a}}{\bar{b}},\frac{\bar{l}_b}{\bar{b}} \right) \\
\bar{\Phi}_{sol}^\succ(b,\infty)&=&\bar{\Phi}_{sol}(\bar{b},\infty,\bar{l}_{sol}) \\
\bar{\Phi}_a^\succ(b,\infty)&=&\bar{\Phi}_{sol} \left(\frac{\bar{b}}{\bar{a}},\infty,\frac{\bar{l}_a}{\bar{a}} \right) \\
\bar{\Phi}_b^\succ(b,\infty)&=&\bar{\Phi}_{sol} \left(1,\infty,\frac{\bar{l}_b}{\bar{b}} \right)
\ea
where the proper form of current should be used in each case and $\bar{\Phi}_{sol}$ on the RHS is to be evaluated using equation \eqref{flux_mean_finite}. In Figure \ref{fig:Figure_5} we show a simple Mathematica code for evaluating the RHS of equation \eqref{flux_mean_finite}.

\begin{figure}[h!]
\centering
\includegraphics[width=\textwidth]{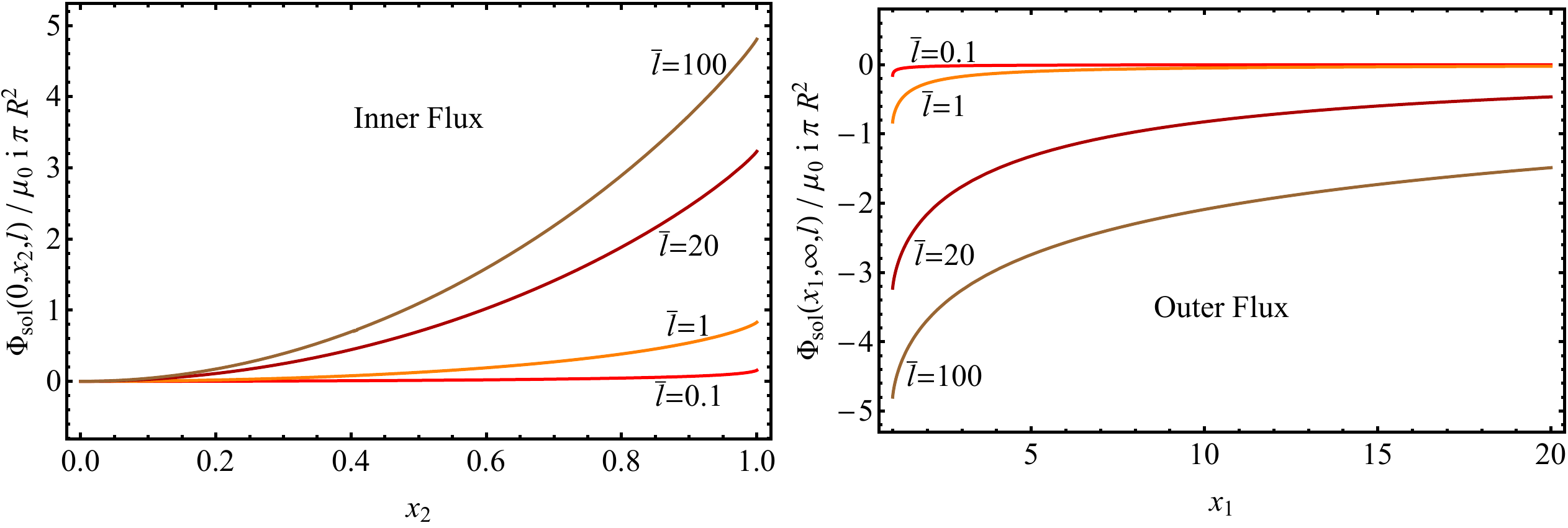}
\caption{Plots of inner and outer fluxes obtained from  \eqref{flux_mean_finite} assuming various values of $\bar{l}$.}
\label{fig:Figure_6}
\end{figure}

In Figure \ref{fig:Figure_6} we show plots of the mean inner and outer fluxes obtained obtained using  \eqref{flux_mean_finite} assuming $\bar{l}=20$. Notice that the outer flux is negative as expected. These plots may be used to obtain the inner and outer fluxes of equations \eqref{tot_lmech} and \eqref{tot_lfield}.

\section{Conclusions}
\label{sec:Conclusions} 

We have presented a general framework for the quantitative demonstration of conservation of total angular momentum in a system that involves both mechanical and field angular momentum. Our framework is based on expressing both the field and the mechanical angular momentum of a system in terms of electric charges and magnetic field fluxes residing on system components.  It is applicable for finite systems and therefore it is more realistic than previous similar analyses. It may be useful in educational laboratories demonstrating the transition from finite realistic system components towards infinite or infinitesimal system components (solenoids).

Our framework extends the applicability of previous quantitative analyses\cite{:/content/aapt/journal/ajp/52/8/10.1119/1.13853, Griffiths} of the FEP by taking into consideration additional physical effects that should be present in any realistic system. In particular we have taken into account the finite size of the solenoid by calculating the effects of the weak magnetic field in the solenoid outer region. In addition, we have taken into account the magnetic field induced by the rotating charge density inside and outside the cylinders. Conservation of total energy momentum was demonstrated in the presence of these additional effects. Our analysis could also be useful in the quantitative understanding of the system considered in the context of a realistic implementation. 

A remaining assumption of our approach is the adiabatic reduction of the current in the solenoid. This allows ignoring electromagnetic radiation. A model in which electromagnetic radiation is taken into account would be an interesting extension of the present study.
\\ \\
\textbf{Numerical Analysis Files}: See Supplemental Material at \href{https://www.dropbox.com/s/267vu6xc6s5rdfh/fdp.zip?dl=0}{here} for the {\it Mathematica file} used for the production of the figures, as well as the figures.

\section*{Acknowledgments}
We would like to thank the referees for their useful suggestions in order to improve the presentation and the quality of the paper.

\raggedleft
\bibliography{fdp}

\end{document}